# Direct evidence of magnetic reconnection onset via the tearing instability

**Mayur R. Bakrania** [1,*], **I. Jonathan Rae** [1,2], **Andrew P. Walsh** [3], **Daniel Verscharen** [1], **Andy W. Smith** [1], **Colin Forsyth**[1], **& Anna Tenerani**[4]

[1]*Department of Space and Climate Physics, Mullard Space Science Laboratory, University College London, Dorking RH5 6NT, UK*

[2]*Department of Mathematics, Physics and Electrical Engineering, University of Northumbria, Newcastle, UK*

[3]*European Space Astronomy Centre, ESA, Urb. Villafranca del Castillo, Villanueva de la Cañada, Madrid, Spain*

[4] *Department of Physics, The University of Texas at Austin, Austin, TX 78712, United States*

Correspondence*:
Mayur R. Bakrania
mayur.bakrania@ucl.ac.uk

## ABSTRACT

Magnetic reconnection is a sporadic process responsible for energy release in space and laboratory plasmas. It is believed that the tearing mode instability may be responsible for the onset of reconnection in the magnetotail. However, due to its elusive nature, there is an absence of in-situ observations of the tearing instability prior to magnetic reconnection in our nearest natural plasma laboratory. Using neural network outlier detection methods in conjunction with Cluster spacecraft data, we find unique electron pitch angle distributions that are consistent with simulation predictions of the tearing instability and the subsequent evolution of plasma electrons and reconnection. We confirm that the events identified via our neural network outlier method are well above the tearing stability threshold based on the criterion detailed in this paper. We find signatures of magnetic reconnection minutes after the majority of tearing observations. Our analysis of the tearing instability provides new insights into the fundamental understanding of the mechanism responsible for reconnection, a process that is ubiquitous in different astrophysical plasma regimes across the universe and in laboratory experiments on Earth.

Keywords: space plasma environments, magnetic reconnection, tearing instability, neural network techniques, outlier detection

## 1 INTRODUCTION

Magnetic reconnection is a universal plasma process that changes the topology of the magnetic field and converts magnetic energy into particle kinetic and thermal energy (Birn and Priest, 2007). This process is responsible for explosive phenomena in laboratory, astrophysical, and space plasmas, such as planetary magnetospheres. The Earth's nightside magnetosphere, i.e. the magnetotail, provides an accessible medium to directly measure reconnection with in-situ spacecraft. Although there are in-situ observations of ongoing magnetic reconnection (Eastwood et al., 2010; Øieroset et al., 2001; Borg et al., 2012; Hwang et al., 2013;

Nagai et al., 2001), the question remains what triggers this process that sporadically changes the magnetic topology, as this process has not been directly measured (Galeev and Zelenii, 1976; Pellat et al., 1991).

The tearing instability is the central candidate mechanism that creates conditions required for the onset of reconnection (Coppi et al., 1966; Chen et al., 1997). In the magnetotail, this instability may occur for sufficiently thin current sheet and small values of the magnetic field normal to the current sheet, and causes quasi-periodic spatial perturbations of the magnetic field and the associated particle distribution functions (Sitnov et al., 2019). The instability takes place within an externally driven current sheet (current sheet thinning caused by an external factor), which usually occurs during the substorm expansion phase (Bessho and Bhattacharjee, 2014). The result of magnetic reconnection is a rearrangement of the magnetic field topology threading the thin current sheet (Galeev and Zelenii, 1976), leading to the formation of plasmoids and X-lines (Zanna et al., 2016), as well as heating and acceleration of the plasma along the field lines. Using kinetic models, two-dimensional Particle-In-Cell (PIC) simulations (Bessho and Bhattacharjee, 2014) have shown that in Earth's magnetotail, the electron tearing instability is the most relevant instability for the initiation of reconnection. However, observational signatures of such a process are still lacking.

On the other hand, PIC simulations have found electron distributions that exhibit a strong counter-streaming field-aligned distribution as the tearing mode develops (Markidis et al., 2012; Zeiler et al., 2001; Buechner and Zelenyi, 1987), followed by an isotropisation across all pitch angles over 5 minutes (Buechner and Zelenyi, 1987). The counter-streaming electron distributions during the early evolution of tearing correspond with a rapid growth in energy and temperature (Buechner and Zelenyi, 1987; Walker et al., 2018). Following this stage, the perpendicular temperature increases (Liu et al., 2014), driving the onset of magnetic reconnection. Here we hypothesise that these field-aligned distributions and their subsequent evolution can be identified in the magnetotail current sheet in order to locate the tearing instability. We apply a neural network outlier detection method to Cluster spacecraft electron data to identify the location where the tearing mode develops. We verify that the plasma and magnetic field parameters at those locations are consistent with tearing instability by applying the stability criterion derived by Schindler et al. (1973); Liu et al. (2014), valid for a collisionless plasma with a finite normal component. It is found that tearing mode events identified via our neural network outlier method are consistent with the tearing instability. Furthermore, data analyses of the subsequent evolution of particles and magnetic fields confirm the typical features found in reconnection simulations. The present work provides strong observational support to theories predicting tearing mode as the onset mechanism for reconnection in the magnetotail.

## 2 METHOD
### 2.1 Data

We use electron data (Laakso et al., 2010) from the Cluster (Escoubet et al., 2001) mission's PEACE (Johnstone et al., 1997; Fazakerley et al., 2010) (Plasma Electron And Current Experiment) instrument on all of the four spacecraft to detect signatures of the tearing mode in the electron distribution functions. Cluster's four spacecraft fly in a tetrahedral formation with a spin period of 4 seconds. We use pitch angle distributions from the PITCH-SPIN data product, which have a 4 s time resolution and are constructed from two instantaneous pitch angle measurements per spin. Each distribution consists of a two-dimensional differential energy flux product with twelve 15° pitch angle bins and 26 logarithmically spaced energy bins ranging from 93 eV to 24 keV. Therefore, each distribution has a dimensionality of 312 (12 × 26). We correct our PEACE measurements with the method presented by Cully et al. (2007) to account for the effect of the spacecraft potential measured by the Cluster-EFW instrument (Gustafsson et al., 2001). We normalise the value of differential energy flux between 0 and 1, based on the value of maximum flux in the dataset, in order to concentrate on the shape of the distribution rather than the flux value, given that the Earth's magnetotail plasma sheet can vary by 5 orders of magnitude with different conditions (Artemyev et al., 2014).

We use the ECLAT database (Boakes et al., 2014) to isolate relevant times for detecting the tearing instability. The ECLAT database uses data from PEACE, FGM (Balogh et al., 1997), and CIS (Rème et al., 2001) to construct a list of plasma regions encountered by the four Cluster spacecraft in the magnetotail from July to October, during the years 2001-2009. The ECLAT database associates the measurement intervals with three magnetotail regions: the plasma sheet, the plasma sheet boundary layer, and the lobes (Hughes, 1995), which are defined by their plasma and magnetic field characteristics. The database also lists times of current sheet crossings at the centre of the plasma sheet. We obtain PEACE data from times when the spacecraft has spent at least 30 minutes in the plasma sheet, as this region is most likely to undergo magnetotail reconnection (Angelopoulos et al., 2008).

## 2.2 Autoencoder

We employ an autoencoder (Hinton and Salakhutdinov, 2006) to detect anomalous distributions from the entire set of plasma sheet data (Bakrania et al., 2020). Autoencoders are a class of unsupervised neural networks which are trained to learn compressed representations of data. These compressed representations are achieved via a 'bottleneck' layer which maps (encodes) the input data to a lower-dimensional latent space, and subsequently reconstructs (decodes) the original input from this latent space. By minimising the reconstruction error, or 'loss' between the input and output data, the autoencoder retains the most important characteristics in the compressed version of the data. For an anomalous distribution, its most important features are not present in the latent space, which results in a large reconstruction error between the input and output data. Autoencoders are therefore an effective method for isolating outliers (Kube et al., 2019). Figure 1 illustrates the typical architecture of an autoencoder (Sakurada and Yairi, 2014). A detailed description of autoencoders is provided by Hinton and Salakhutdinov (Hinton and Salakhutdinov, 2006).

We construct our autoencoder using the Keras library (Chollet et al., 2015). Building an autoencoder requires the definition of the number of neurons in each layer. The number of neurons in the input and output layers equals the dimensionality of each distribution: 312 in our case. We set the number of neurons in the bottleneck layer at 32, representing a compression factor of 9.75. Each layer uses an activation function to pass on signals to the next layer (Kube et al., 2019). For the encoder part, we use the ReLU activation function (Hahnioser et al., 2000), and for the decoder part, we use the sigmoid activation function (Chandra and Singh, 2004), which normalises the output between 0 and 1. We then define the loss function and optimiser, which the autoencoder uses to representatively compress and reconstruct the input data. We choose the binary cross-entropy loss function (de Boer et al., 2005) and the Adadelta (Zeiler, 2012) optimiser. All activation functions, loss functions, and optimisers are available in the Keras library. We set the number of epochs to 500 and the batch size, i.e. the number of distributions propagated through the network at each epoch, to 256. The validation split ratio defines the ratio of distributions that remain 'unseen' to the autoencoder in order to avoid overfitting. We set this to 1/12. At each epoch a training loss and validation loss value are produced, which both converge to <0.1 after training, indicating that the autoencoder accurately reconstructs the majority of the dataset.

To isolate the field-aligned tearing distributions from the dataset, we calculate the mean square error (MSE) between each original and reconstructed distribution. We set the MSE threshold to 99.5%, which locates all distributions which have a MSE in the upper 0.5% of the dataset. We subsequently visually inspect each anomalous distribution to find signatures of the tearing instability based on the flux anisotropy, as outlined in section 1. We then obtain a list of outlier measurements to be studied further.

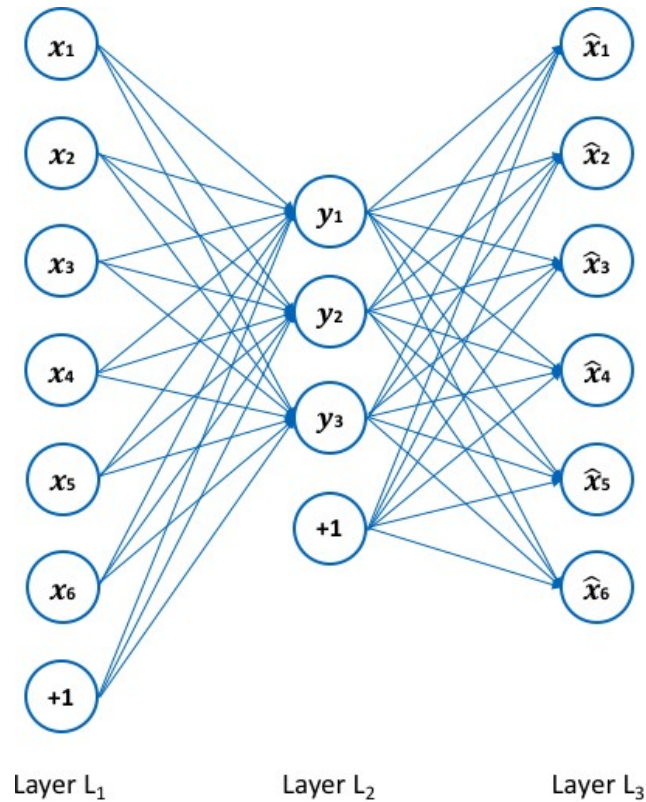

**Figure 1.** The architecture of an autoencoder, adapted from Sakurada and Yairi (2014). Layer $L_1$ represents the input data, with each circle representing a neuron corresponding to a measurement bin, layer $L_2$ is the encoded data in latent space, and layer $L_3$ is the reconstructed data. The circles labelled '+1' are 'bias units', which are parameters the autoencoder adjusts during training to improve its performance.

### 2.3 Coordinate system

For the intervals which contain a current sheet crossing, we transform the coordinate system from the GSM coordinate system to the local ($LMN$) coordinate system. This coordinate system provides a more accurate representation of the magnetic field vectors as it takes into account the current sheet tilt. We obtain the local coordinate system from a minimum variance analysis (Sonnerup and Scheible, 1998) of the magnetic field data during a short interval before each instance of tearing. In this coordinate system, $L$ is in the direction of the anti-parallel magnetic field, $M$ is in the direction of the current, and $N$ is in the normal direction to the current sheet. Our algorithm identifies an event on 07/08/2004. For this event, we obtain this coordinate system from the minimum variance analysis of the magnetic field data from 23:01:02 UT to 23:06:00 UT. The vector representations of the local magnetic coordinates are: $L$ = (0.9859, -0.1110, -0.1253), $M$ = (0.1491, 0.9221, 0.3570) and $N$ = (0.0759, -0.3707, 0.9257) in GSM coordinates. We also account for a small out-of-plane ($M$-direction) guide field of +1.06 nT during this event.

### 2.4 Tearing mode stability criteria

We determine whether an outlier event is tearing unstable based on the instability criteria set out by Schindler et al. (1973), Liu et al. (2014), and in other studies (see below). Simulations (Bessho and Bhattacharjee, 2014) show that the electron tearing instability only occurs when the magnitude of $B_N$ is small (<10 nT), in agreement with other studies (Galeev and Zelenii, 1976; Pellat et al., 1991). Schindler et al. (1973) predicts that $B_N$ must be positive for the instability to arise. Both theory (Coppi et al., 1966; Zanna et al., 2016; Schindler et al., 1973) and simulations (Bessho and Bhattacharjee, 2014), show that current sheet thinning is especially important for the tearing mode instability.

By applying the energy principle to two-dimensional Vlasov equilibria, Schindler et al. (1973) derived a stability criterion that predicts instability when $k\rho_e \gtrsim 1$, where $k$ is the wavenumber of the perturbation and $\rho_e$ the electron gyroradius associated with the normal magnetic field component. This criterion can be equivalently written as (Liu et al., 2014):

$$b\frac{L_N}{d_i} < \frac{f}{\zeta}\sqrt{\frac{m_e T_e}{m_i T}}, \tag{1}$$

where $b = B_n/B_0$, $B_0$ is the lobe magnetic field, $L_N$ is the half-current sheet height, $d_i$ is the ion inertial length, $f = k_L L_N$, $\zeta$ is a parameter which we set to 1, $T = T_e + T_i$, and $T_e$ and $T_i$ are the electron and ion temperatures.

After identifying possible candidate measurement points for tearing, we check if each candidate matches the theory and the simulations (see equation 1) by comparing the corresponding magnetic field measurements (GSM coordinate system) from the FGM instrument, and electron temperature measurements from the PEACE instrument. This includes checking if there is a rapid growth in temperature, in conjunction with an isotropisation, shortly after the anomalous distribution is observed, in line with the results from PIC simulations described in section 1 (Buechner and Zelenyi, 1987; Liu et al., 2014; Walker et al., 2018). We also check if the $B_N$-field exhibits a small positive component at the time of the outlier distribution. If the spacecraft crosses the current sheet near to the tearing distribution observation, we also determine whether the normal component of the magnetic field at the crossing is smaller or equal to the value of the normal component during tearing, in compliance with equation (1).

## 2.5 Hall quadrupole field

To link our events to magnetic reconnection, we look for evidence of a Hall quadrupole magnetic field and flow reversals. The Hall quadrupole field shows that $B_M$ exhibits a correlation with $v_L$ above the current sheet (Northern Hemisphere), and an anti-correlation with $v_L$ below the current sheet (Southern Hemisphere). During the intervals in which we observe flow reversals, a small out-of-plane ($M$-direction) guide magnetic field (Denton et al., 2016) may be present which needs to be accounted for. Averaging $B_M$ prior to the flow reversal informs us of the magnitude of this guide field. We therefore correct for this by shifting all $B_M$ values in this interval until the average $B_M$ vanishes.

## 3 RESULTS
## 3.1 Tearing mode stability check

In order to find signatures of the tearing instability, we apply our neural network outlier detection method (Bakrania et al., 2020) to Cluster-PEACE data from the magnetotail plasma sheet (see section 2.2). From the collection of outlier events, we identify 15 separate time intervals, between 2001 and 2009, (see Table 1) that all fulfill the criteria for field-aligned distributions described in section 1, as well as the magnetic field criteria (see section 2.4). We refer to the field-aligned distributions as tearing distributions. Figure 2 shows the results of our stability analysis for all events. We create the plot by inputting measurements during each event into equation (1) (Liu et al., 2014), with the only unknown being $f$. The figure 2 confirms that all events adhere to the instability criterion in equation (1). Using multi-spacecraft measurements, we calculate for this figure the magnetic field, current sheet height (see below), ion internal length, mass ratio and temperature ratio for each event. As the parameter *f* relates to the wavenumber, we provide different estimates for *f* based on the analysis by Liu et al. (2014). For a realistic mass ratio ($m_e/m_i$) of 1/1836, Liu et al. (2014) predict that *f* = 0.91. Figure 2 shows that all events are below the *f* = 0.91 line and therefore adhere to the instability criterion.

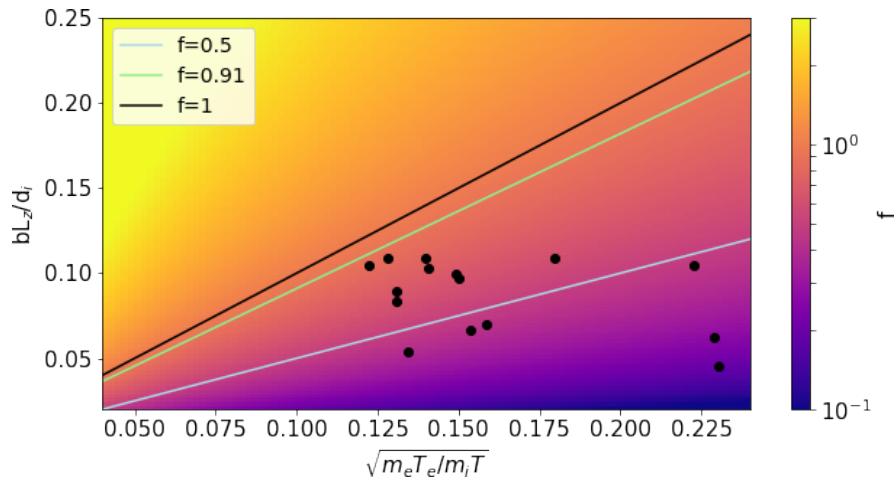

**Figure 2. Verification of the tearing instability.** The black dots represent each of the 15 events identified by our neural network method. The axes correspond to the terms on the left and right hand side of equation (1), while the background colour represents the parameter *f* in equation (1). The straight lines represent the boundaries for different values of *f*, below which the plasma is tearing unstable (Liu et al., 2014).

## 3.2 Case study

In this section we discuss the event which occurred on 07/08/2004. We discuss the rest of the events in section 3.3. In this case study event, Cluster passed from North to South through the magnetotail, crossing the current sheet at 23:27:13 UT before encountering an electron population that shows signatures of the tearing mode instability, followed by a crossing through a diffusion region. The first outlier distribution, or tearing distribution, was detected by the C4 spacecraft at 23:29:05 UT, when the spacecraft was below the current sheet in the position (-16.4, -9.0, 0.1) $R_E$ (where 1 $R_E$ = 6,371 km) in GSM (Geocentric Solar Magnetospheric) coordinates. The remaining three Cluster spacecraft, which were separated by 0.2 $R_E$ from the C4 spacecraft, also observed similar tearing distributions.

Figure 3 shows how the electron temperature, magnetic field, and plasma flow evolve during this particular event. The times of tearing distributions are indicated by red dots. The lines A, B, and C represent the times of the distributions which we show in Figure 4, line A represents a time before the detection of the tearing distribution at which C4 crossed the tail current sheet, line B represents the time of a tearing distribution detection, and line C represents a time after this detection.

According to the data presented in Figure 3, the tearing mode occurs within the unstable magnetic field regime (Schindler et al., 1973; Liu et al., 2014) (see section 2.4). At the first indication of a tearing distribution (red dot), the normal magnetic field, $B_N$, component has a value of ∼5 nT. This value is larger than the value of $B_N$ at time A by a factor of ∼10 during a current sheet crossing, in line with the instability criterion in equation (1) (see section 2.4). We also plot the $B_N/B_{N=0}$ ratio to show how this criterion is met during the period when the tearing distributions are observed. The temperature increases rapidly after the detection of the tearing distribution, in conjunction with an increase in electron anisotropy ($T_\perp/T_\parallel$) from

**Table 1.** Times of the 15 tearing events, along with the spacecraft which detected the tearing distributions and the subsequent magnetic reconnection signatures.

| Event # | Spacecraft | Time of Tearing | Reconnection Seen |
|---|---|---|---|
| 1 | C1, C2, C3, C4 | 17:11:12 UT, 18/08/2002 | Yes |
| 2 | C1, C2, C3, C4 | 13:26:57 UT, 18/09/2002 | Yes |
| 3 | C1, C2, C3, C4 | 20:48:37 UT, 02/10/2002 | Yes |
| 4 | C1, C2, C3, C4 | 21:21:21 UT, 02/10/2002 | Yes |
| 5 | C1, C2, C3, C4 | 16:43:17 UT, 17/08/2003 | Yes |
| 6 | C1, C2, C3, C4 | 18:57:27 UT, 24/08/2003 | Yes |
| 7 | C1, C2, C3, C4 | 06:19:12 UT, 04/10/2003 | Yes |
| 8 | C1, C2, C3, C4 | 23:29:05 UT, 07/08/2004 | Yes |
| 9 | C2, C3, C4 | 02:16:57 UT, 10/08/2005 | Yes |
| 10 | C2, C3, C4 | 02:58:28 UT, 10/08/2005 | No |
| 11 | C3, C4 | 21:52:25 UT, 07/08/2008 | No |
| 12 | C1 | 01:19:21 UT, 15/09/2008 | Yes |
| 13 | C1, C3, C4 | 01:17:59 UT, 02/09/2009 | No |
| 14 | C1, C3, C4 | 21:19:37 UT, 13/09/2009 | No |
| 15 | C1, C3, C4 | 21:31:52 UT, 13/09/2009 | No |

∼0.6 to 1. These anisotropy changes agree with PIC simulations (Buechner and Zelenyi, 1987; Liu et al., 2014; Walker et al., 2018) that predict that the tearing distribution evolves into an isotropic distribution during the growth of the tearing instability. Figure 4 illustrates this temperature and anisotropy evolution. A low energy isotropic distribution prior to tearing evolves into a strong field-aligned distribution, followed by an isotropic distribution with a higher temperature, as expected from theory.

After noting the close correspondence between the observations and simulations of the tearing mode, we can gain further insight into the time evolution of the tearing mode. Firstly, the field-aligned population lasts for just over two minutes, during which the total temperature remains approximately stable at 400 eV. After the last tearing distribution is detected, a rapid growth in temperature reaches a peak of 1979 eV, 336 s after the first tearing distribution. As the tearing mechanism is characterised by the evolution of a field-aligned distribution into an isotropic distribution at higher temperatures, we attribute this time of 336 s to the growth time of the instability for this event. After this temperature increase, the spacecraft observes direct evidence of magnetic reconnection at 23:36:52 UT, i.e. 467 s (or 7 min 47 s) after the first tearing distribution, as shown by the simultaneous reversal in $B_N$ and $v_L$ (from positive to negative values) in Figure 3 after the tearing distributions.

Nagai et al. (2001) find that the reconnection process generates a current system known as the Hall quadrupole magnetic field (Karimabadi et al., 2004). Figure 5 shows the out-of-plane magnetic field ($B_M$) values in the along-current sheet ($B_L$-$v_L$) plane after our 07/08/2004 tearing event, which confirms the presence of this characteristic Hall quadrupolar field (see section 2.5). The scatter of the points in Figure 5 shows that negative $B_M$ values dominate in the upper left and bottom right quadrants, while positive $B_M$ values dominate in the other two quadrants. This pattern corresponds to the quadrupole signature of correlation (anti-correlation) with $v_L$, i.e. the speed of plasma flow in the *L*-direction, in the Northern (Southern) Hemisphere (this quadrupole pattern is also illustrated in Figure 6). The bar chart in Figure 5b confirms this quadrupole observation, as there is a clear dominance of positive or negative $B_M$ depending on the respective quadrant, in keeping with the expected signatures from magnetic reconnection.

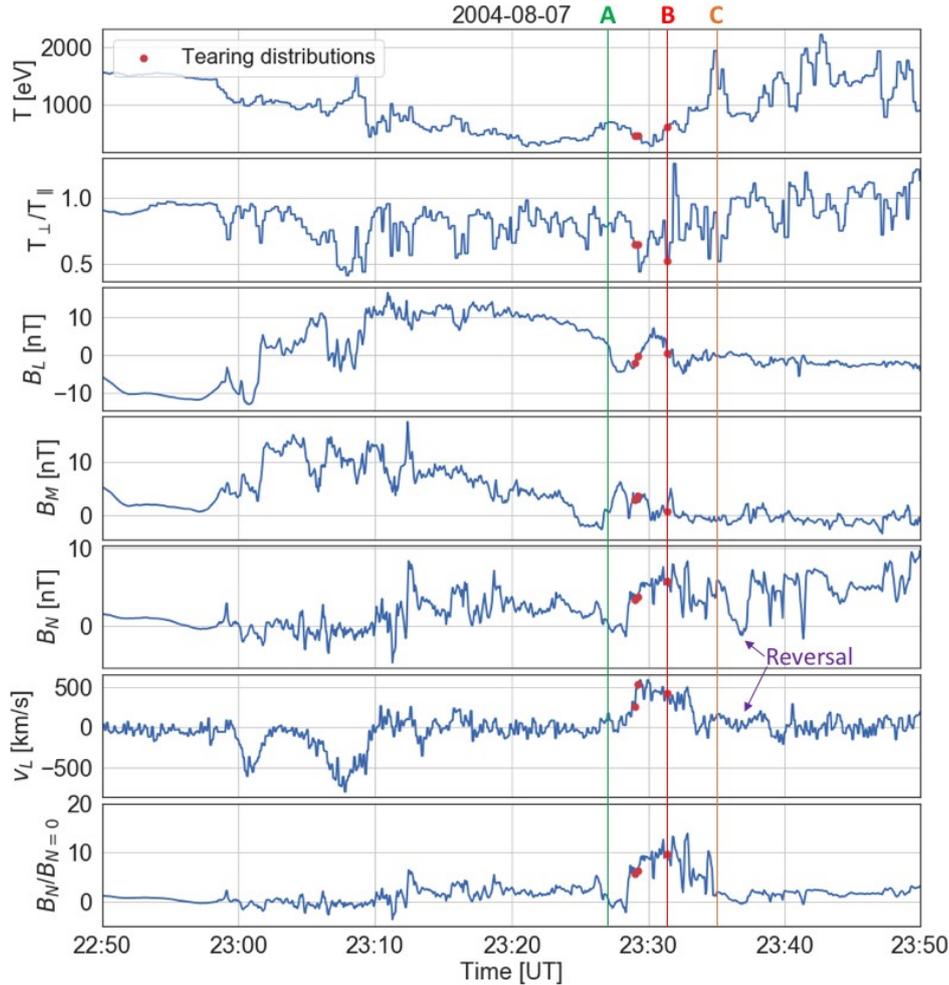

**Figure 3. Spacecraft measurements indicating the presence of the tearing instability.** The plasma and magnetic field parameters obtained by the C4 spacecraft across times 22:50:00 - 23:50:00 on 07/08/2004. From top to bottom: electron temperature, electron $T_\perp/T_\parallel$, magnetic field in the *L, M, N*-directions respectively (in local magnetic coordinates), plasma flow velocity in the *L*-direction, and the $B_N/B_{N=0}$ ratio. A flow reversal is highlighted in the $B_N$ and $v_L$ panels. The red points indicate times of the tearing distributions identified by our outlier detection method. The lines labelled A, B, and C represent times of pre-tearing, tearing, and post-tearing distributions respectively. We show the distributions measured at these times in Figure 4.

Figure 6 provides a schematic detailing the evolution of a laminar current sheet into a reconnection site via the tearing instability consistent with the observations of our 07/08/2004 event. We also illustrate the magnetic islands (Ishizawa and Nakajima, 2010) in 3D along with their expected location in near-Earth space. Each panel corresponds to the timestamps A, B, or C in Figure 3. The green crosses show the C4 spacecraft position at each time, and the green arrow shows the overall trajectory, as informed by the magnetic field measurements. The background colours illustrate the electron temperature in the reconnection site and the surrounding magnetic islands (Lu et al., 2019), with the scale informed by temperature measurements in our event. As the diagram shows, the magnetic island formation coincides with an increase in electron temperature, as observed by the spacecraft after time B in Figure 3. The spacecraft then observes the X-point, as signified by the flow reversal in Figure 3, which aligns with a region of lower temperature. Subsequently, the spacecraft observes the second region of high temperature.

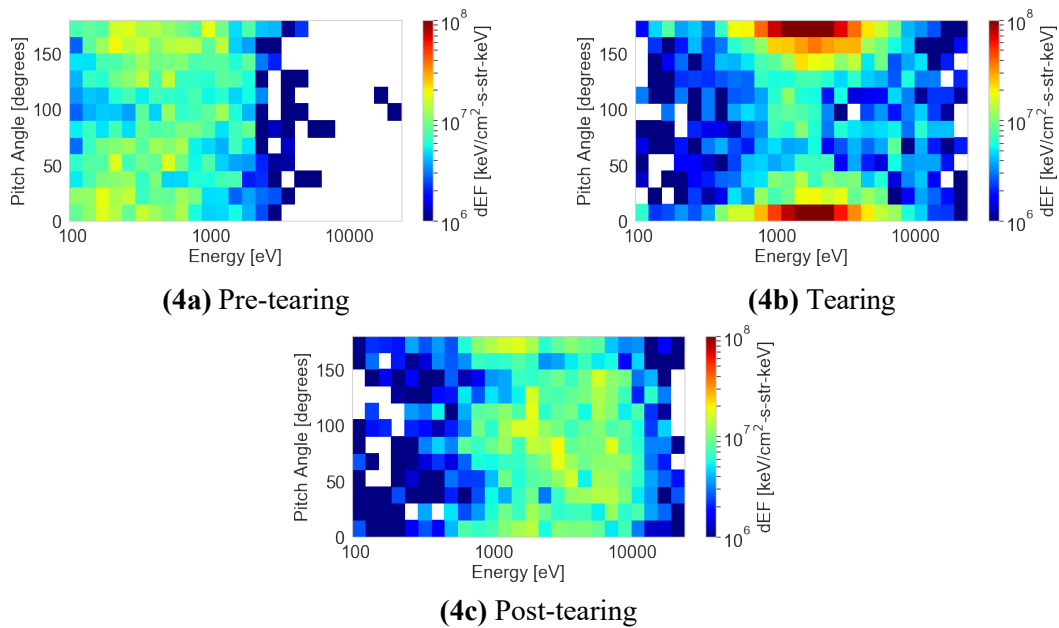

**(4a)** Pre-tearing      **(4b)** Tearing

**(4c)** Post-tearing

**Figure 4. The evolution of a tearing distribution into an isotropic distribution with a higher bulk energy.** The differential energy flux distributions as a function of pitch angle and energy, which correspond to the times A (23:27:13 UT), B (23:34:23 UT), and C (23:34:41 UT) in Figure 3 as measured by the C4 PEACE instrument.

From our case study, we construct a picture of the tearing instability, incorporating the temperature profiles and electron distribution functions of a tearing unstable plasma. We also relate the instability to the onset of reconnection. Our statistical survey will allow us to further quantify the tearing instability and its relationship with reconnection, building in analysis from all 15 of our tearing events.

### 3.3 Statistical survey of tearing events

Our method finds 14 other events that we use to test for the consistency with the tearing mode. We carry out a superposed epoch analysis of the SML (SuperMAG Auroral Lower) (Gjerloev, 2012) indices around each of the 15 tearing events, to determine how they relate to substorm phase (Forsyth et al., 2015). At the end of the substorm expansion phase, the SML index is at its maximum magnitude, and subsequently decreases in the recovery phase. Figure 7 shows that the tearing events occur both during the expansion and recovery phase, with the average time located at the end of the expansion phase. As the growth phase is by far the most prevalent phase of a substorm (Forsyth et al., 2015), a chi-squared test (Tallarida and Murray, 1987) between the tearing events and plasma sheet distributions shows that the occurrence of all tearing events during the expansion and recovery phase is statistically significant to a p-value of $<<0.05$, as the two distributions cannot be derived from the same source. This analysis confirms that our tearing events occur under an externally driven current sheet (which is the case during the expansion phase), as expected from simulations (Bessho and Bhattacharjee, 2014).

Our calculations of the current sheet thickness (Thompson et al., 2005) using equation (2) also reveal evidence of current sheet thinning, further showing that the current sheet is being externally driven by the substorm expansion phase. Saito (2015) find that the average current sheet thickness is roughly 6 $R_E$. Across each of the tearing events, we calculate an average thickness of 0.6 $R_E$, with a minimum value of 0.39 $R_E$ and a maximum value of 0.77 $R_E$.

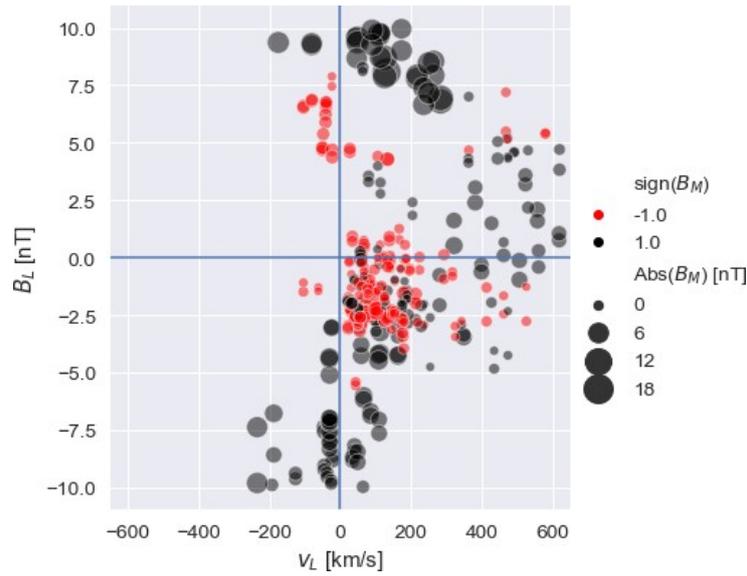

(5a)

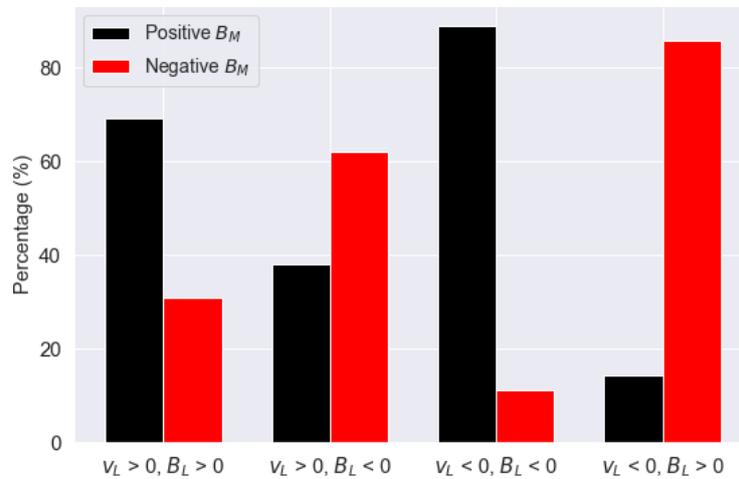

(5b)

**Figure 5. (a) Statistical determination of a reconnection site.** Quadrupole plot showing $B_M$ as a function of $B_L$ and $v_L$, across a 10 minute window centered on the $B_N$ reversal in Figure 3 (at 23:36:52 UT) after the tearing instability distributions. Black dots correspond to $B_M > 0$ and red dots correspond to $B_M < 0$. The size of the dots is proportional to the magnitude of $B_M$. **(b)** The percentage of instances with $B_M > 0$ and $B_M < 0$ in each quadrant.

In our statistical survey, we determine the growth times of the instability across all events, based on the time between the first detection of a tearing distribution and the subsequent peak in temperature. In the 15 events, the average growth time is 601 s, with a lower quartile of 348 s and an upper quartile of 766 s. The minimum and maximum growth times are 224 s and 1200 s respectively. The average peak temperature across all events is 2691 eV, with a lower quartile of 1766 eV and an upper quartile of 2871 eV. The average change in temperature is 1712 eV. Our analysis shows large variations in the temperature of the tearing unstable plasma. As expected (Galeev and Zelenii, 1976), we do not observe a correlation between the temperature and growth time of the instability.

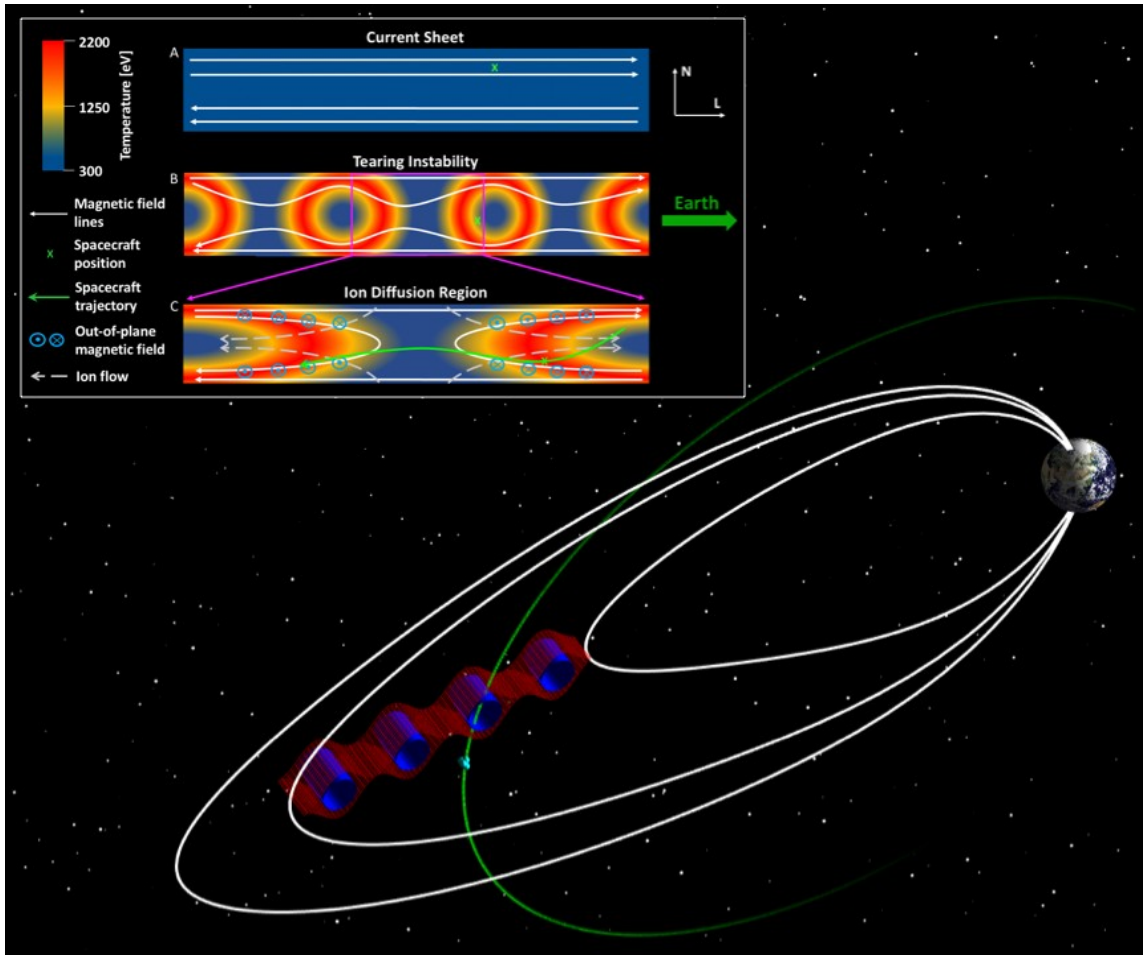

**Figure 6. Illustration of the evolution of magnetic field lines from a flat current sheet into a reconnection site via the tearing instability, with a 3D representation of the magnetic islands in the Earth's vicinity.** The background colours represent the local electron temperature (Lu et al., 2019), with the scales based on the temperatures observed by the C4 spacecraft during the 07/08/2004 tearing event. The white lines represent the magnetic field lines. The three panels correspond to the times A, B, and C in Figure 3, and the green crosses show the location the C4 spacecraft at each time. The green arrow shows the overall trajectory of the spacecraft across this reconnection region in our case study event. The Hall quadrupole field (blue dots and crosses) and corresponding ion flows (dotted grey arrows) are also displayed.

Using multi-spacecraft techniques by cross-referencing observations of tearing distributions with the positions of each spacecraft, we also constrain the size of the region undergoing tearing. In the 07/08/2004 event (Figure 3), we find that all four Cluster spacecraft observe the characteristic tearing distributions. The maximum distance between the spacecraft is 0.20 $R_E$ at this time, and for all events, when the spacecraft are less than 1 $R_E$ from an observed tearing distribution, a similar tearing distribution is observed; spacecraft that are more than 1 $R_E$ from an observed tearing distribution do not observe tearing distributions. However, all four spacecraft in all tearing events observe the characteristic temperature rise after the tearing distribution, as illustrated by the red bands in figure 6, suggesting the region of subsequent particle energisation from tearing-initiated reconnection extends to at least 1.36 $R_E$ in the surrounding plasma (the largest distance between two spacecraft across all tearing events).

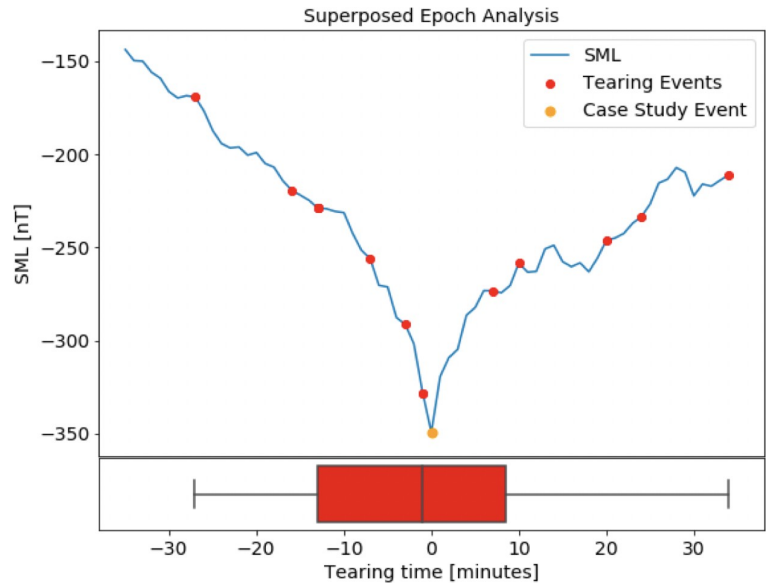

**Figure 7. Superposed epoch analysis of SML relating tearing to substorms.** The spread of tearing times for all 15 events compared to the superposed epoch analysis of SML measurements across each of the 15 events. The blue curve is calculated by averaging the SML peak around each of the 15 tearing events, while the location of the red dots represents the relative timings of each particular tearing event in comparison to the its associated peak in SML. The box plot underneath summarises the distribution of tearing event timings in comparison to the peak in SML. The vertical lines of the red box show the lower quartile, mean, and upper quartile of the relative times, while the two longest lines show the earliest and latest tearing event.

Of the 15 tearing events, we observe signatures of reconnection after 10 of them. For 8 of these 10 events, the spacecraft crosses the current sheet and we observe the characteristic Hall quadrupole magnetic field (Borg et al., 2012). Eastwood et al. (2010) also confirm the presence of a reconnection site for 7 of these events. For the remaining two reconnection events, the spacecraft do not traverse the current sheet; rather they remain below the current sheet and therefore cannot detect all quadrants of the Hall quadrupole field. In these two events, the spacecraft and current sheet are moving closer together, so we observe an anti-correlation between $B_M$ and $v_L$, i.e. two of the four expected quadrants, highlighting the presence of a diffusion region associated with reconnection (Eastwood et al., 2010). For the other 5 tearing events where the spacecraft do not observe a diffusion region, the spacecraft remain in the Southern Hemisphere and the $B_L$ field increases in magnitude, indicating that the spacecraft and current sheet are moving apart.

Figure 8 shows a summary of the quadrupolar signatures expected from magnetic reconnection for all 10 of the tearing-reconnection events. It can be seen from Figure 8 that there is a clear average quadrupole signature which the corresponding bar chart confirms. There are similar time delays between the first tearing distribution and the observation of a reconnection X-point across all 8 events where a reconnection X-point is observed. This delay varies between 5 min 32 s and 8 min 20 s, a variation possibly caused by unpredictable current sheet flapping (Sergeev et al., 2003; Forsyth et al., 2009). These findings provide an insight into the characteristic timescale (of a few minutes) of reconnection site formation as a result of the tearing instability.

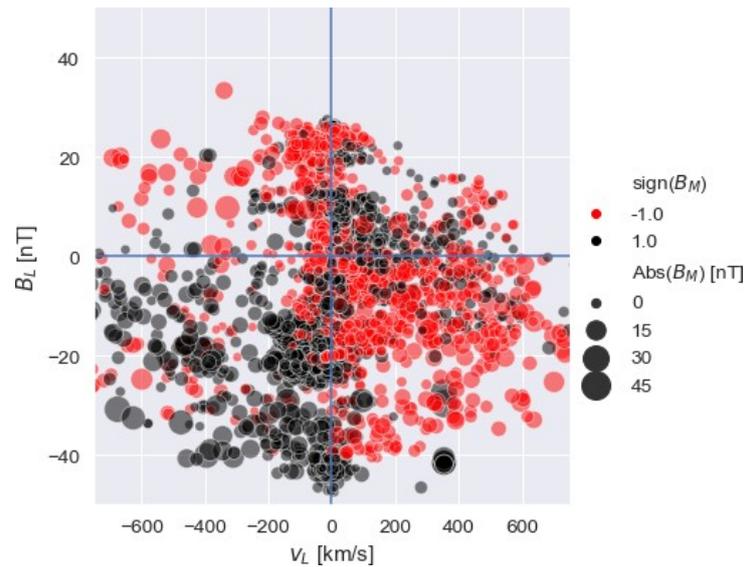

(8a)

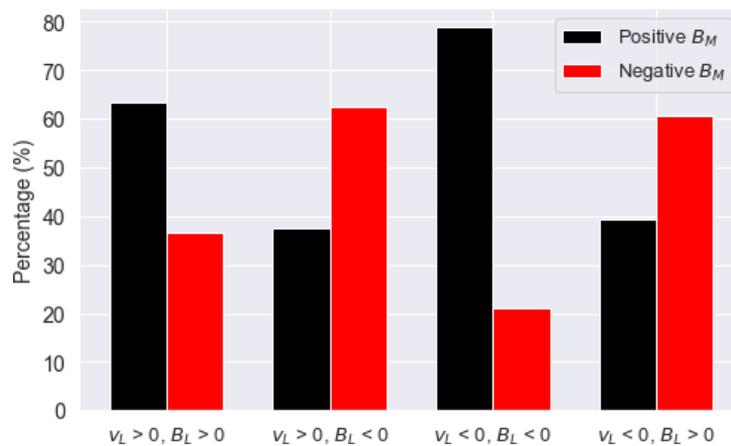

(8b)

**Figure 8.** **(a)** Quadrupole plot showing the out-of-plane magnetic field $B_M$ in the $B_L$-$v_L$ plane, for 10 of our observations where a reconnection signature is observed after the tearing instability distributions. The features of this plot are consistent with those seen in Figure 5. **(b)** The percentage of instances with $B_M > 0$ and $B_M < 0$ in each quadrant.

## 4 DISCUSSION

The tearing mode instability is the dominant mechanism responsible for the onset of magnetic reconnection in high-$\beta$ plasmas (Galeev and Zelenii, 1976). In this study, we present the first in-situ observations of distributions that are consistent with the tearing mode instability in the high-$\beta$ environment of Earth's magnetotail plasma sheet. These tearing mode events are discovered by applying a neural network outlier detection method (Bakrania et al., 2020) to Cluster-PEACE data of the electron velocity distribution functions taken within the plasma sheet. These neural network methods have applications in any plasma environment where in-situ measurements are available. We identify 15 separate cases of the tearing mode, and confirm that they fulfill the theoretical tearing instability criteria (Schindler et al., 1973) and match simulations of the evolution of electron anisotropy and temperature.

As magnetotail current sheet plasmas share similarities with other environments in which the tearing instability is important, our observations provide a fundamental understanding of the mechanism responsible for the initiation of magnetic reconnection. We detect signatures of magnetic reconnection after 10 of the 15 events, producing observational evidence for the link between tearing and reconnection, a link which has only been shown in simulations to date (Bhattacharjee et al., 2009). The lack of reconnection signatures after the other 5 tearing events can be explained by the relative position and trajectory of the spacecraft with respect to the reconnecting current sheet (based on the increase in magnitude of $B_L$).

Our observations enable us to make the first experimental measurements of the growth times of the tearing mode, improving our understanding of the timescales of the instability. The growth time of each tearing event varies between 4 and 20 minutes, and with no apparent correlation between the tearing growth time and the value of the subsequent peak temperature, which is a measure of the energisation of the plasma following reconnection. In order to understand how the growth of the tearing instability relates to the properties of the surrounding plasma, we calculate the characteristic Alfvén timescales ($\tau_A$) in 6 of the system for each event. We find that the ratio between the tearing and Alfvén timescales is $\sim$100 across all 15 events. These results show that the relative growth rate of the tearing mode is in fact one order of magnitude lower than predicted in simulations (Walker et al., 2018; Wang et al., 1988).

The main source of uncertainty in our estimation of the growth time lies in the assumption that the spacecraft observe the true start and end of tearing, which cannot be unambiguously determined by in-situ observations alone. If the spacecraft are not observing the true start of the tearing instability, then our calculated time delays may actually characterise the physical distance between the tearing distribution and isotropic distribution, rather than the growth time. As our calculations of the growth rate are underestimates of the true growth rate, the discrepancy between our calculations and previous studies, regarding the growth time to Alfvén timescales, may not be as large.

With the aid of the SOPHIE dataset, we can link the occurrence of the tearing events, to the substorm phase at the times to our 15 tearing events. We find that the tearing instability is significantly more likely to occur during the expansion and recovery phase, than the growth phase, and on average the tearing instability occurs on the boundary between the expansion and recovery phase. As the expansion phase is initiated due to an externally driven current sheet, we confirm that this externally driven current sheet is also a prerequisite for tearing in the magnetotail, as expected from simulations.

The time delay between the first tearing distribution and the formation of the reconnection site varies within a narrow range of 5 min 32 s to 8 min 20 s across our events. This narrow range of times points towards a consistent delay between tearing and reconnection site formation, similar to that found in simulations (Chen et al., 1997). Our multi-spacecraft analysis shows that the size of the tearing region itself is less than 4 $R_E$. More high cadence multi-spacecraft measurements, such as measurements from the MMS (Sharma and Curtis, 2005) (Magnetospheric Multiscale Mission) mission, will be important in furthering this investigation into the size and location of tearing regions.

## 5 CONCLUSION

Magnetic reconnection is a ubiquitous process in space, laboratory, and astrophysical plasmas that converts magnetic energy to kinetic energy and results in a large variety of energetic events, including aurora, solar flares, astrophysical jets, and tokamak disruptions. Since the start of the space age, reconnection has been widely studied in different physical environments. Relatively little has been known, however, about the process by which reconnection is triggered. Until now, theory and simulations have provided most of the insight into this elusive kinetic plasmoid instability mechanism. The novel machine learning techniques we employ make our identification of the tearing instability possible on a consistent basis. In 10 years of high-resolution magnetotail plasma sheet crossings, we find 15 clear examples of tearing unstable distributions that occur prior to observations of magnetic reconnection, providing a clear link between these two physical processes. We show that the time delay between the two processes is of the order of a few minutes, which is also similar to the growth time of the instability prior to reconnection. Furthermore, we show that the timing of these events are linked to current sheet thinning, which occurs during the substorm expansion and recovery phases, confirming that the electron tearing instability results from externally driven processes.

Our study serves as the groundwork for future studies that would investigate how tearing mode growth times vary across different plasma environments, from the magnetosphere to the solar corona and beyond. We provide a comprehensive analysis of the temperature profiles and timescales during the tearing instability, taking a significant step in solving the longstanding problem of reconnection initiation with real-world data.

## CONFLICT OF INTEREST STATEMENT

The authors declare that the research was conducted in the absence of any commercial or financial relationships that could be construed as a potential conflict of interest.

## AUTHOR CONTRIBUTIONS

MRB developed the method described in the manuscript, tested it on the magnetotail data and wrote the manuscript. IJR was the lead supervisor who guided the direction of the project and provided insight at every stage. APW provided expertise on the magnetotail and the various populations that we observed, aiding the evaluation of our method. DV was also important in classifying the plasma regimes and provided insights into the physical processes governing electrons in space plasmas. AWS was key to the development of the method due to his expertise in machine learning. CF provided assistance with figure preparation. AT provided key insights into the tearing instability. All co-authors made important contributions to the manuscript.

## FUNDING

MRB is supported by a UCL Impact Studentship, joint funded by the ESA NPI programme. IJR the STFC Consolidated Grant ST/S000240/1 and the NERC grants NE/P017150/1, NE/P017185/1, NE/V002554/1, and NE/V002724/1. DV is supported by the STFC Consolidated Grant ST/S000240/1 and the STFC Ernest Rutherford Fellowship ST/P003826/1. AWS is supported by the STFC Consolidated Grant ST/S000240/1 and by NERC grants NE/P017150/1 and NE/V002724/1. CF is supported by NERC Independent Research Fellowship NE/N014480/1, NERC grants NE/V002724/1, NE/V002554/2, NE/P017185/2 and NE/P017150/1 and STFC Consolidated Grant ST/S000240/1.

## ACKNOWLEDGMENTS

We thank the Cluster instrument teams (PEACE, FGM, CIS, EFW) for the data used in this study, in particular the PEACE operations team at the Mullard Space Science Laboratory. We also acknowledge the European Union Framework 7 Programme, the ECLAT Project FP7 Grant no. 263325, and the ESA Cluster Science Archive. We also thank Clare Watt and Sarah Matthews for their advice on this manuscript.

## DATA AVAILABILITY STATEMENT

The datasets analysed in this study can be found in the Cluster Science Archive (https://csa.esac.esa.int/csa-web/).

# 6 APPENDIX A

To calculate the current sheet thickness, $a$ (where $a = 2L_N$), we use the following equation (Thompson et al., 2005):

$$a = 2 \frac{B_0^2 - B_L^2}{\mu_0 B_0 J_M}, \tag{2}$$

where $B_0$ is the lobe magnetic field strength, obtained from Bakrania et al. (2020), $B_L$ is the local magnetic field in the $L$ direction (depending on the coordinate system), and $J_M$ is the local current density in the $M$ direction. We obtain the current density using multi-spacecraft measurements from (Perri et al., 2017; Dunlop et al., 1988):

$$\mu_0 \mathbf{J}_{ijk} \cdot (\Delta \mathbf{r}_{ik} \times \Delta \mathbf{r}_{jk}) = \Delta \mathbf{B}_{ik} \cdot \Delta \mathbf{r}_{jk} - \Delta \mathbf{B}_{jk} \cdot \Delta \mathbf{r}_{ik}, \tag{3}$$

where $i, j, k$ represent the C1, C3, and C4 spacecraft respectively. $\mathbf{J}_{ijk}$ and $\mathbf{B}_{ijk}$ are the local current density and magnetic field. For our 07/08/2004 event, we estimate a current sheet width of 0.72 $R_E$, corresponding to ~2 ion gyroradii.

The local Alfvén speed is calculated as

$$v_A = B_L / \sqrt{4\pi\rho_0}, \tag{4}$$

where $B_L$ is the $L$-component of the background magnetic field and $\rho_0$ is the plasma density. The magnetic field and ion density used for calculating the Alfvén speed (Eq. 4) are obtained by averaging the measurements across all Cluster spacecraft when they were outside the current sheet, within 30 min of the tearing event. With an average background density and $B_L$ field of 0.14 cm$^{-3}$ and 13.1 nT respectively, we calculate an Alfvén speed of $7.6 \times 10^5$ m/s for our case study event.